\documentclass[letterpaper]{article} 
\usepackage{aaai2026}                
\usepackage{times}                   
\usepackage{helvet}                  
\usepackage{courier}                 
\usepackage[hyphens]{url}            
\usepackage{comment}
\usepackage{graphicx}                
\usepackage{natbib}                  
\usepackage{caption}                 
\usepackage[utf8]{inputenc}          
 
\usepackage{booktabs}
\usepackage{tabularx}
\usepackage{enumitem}
\usepackage{siunitx}
\usepackage{makecell}
\usepackage{ifthen}
 
\newboolean{showChanges}
\setboolean{showChanges}{false}

\newcommand{\oldtext}[1]{}

\newcommand{\deletes}[1]{}
\newcommand{\qian}[1]{}
\newcommand{\jenny}[1]{}
\newcommand{\malte}[1]{}
\newcommand{\InLineToDo}[1]{}
 
\title{From Fair Representation to Just Recognition in Generative AI}
 
\author{
    Severin Engelmann \&
    Daniel Susser
}
\affiliations{
    Department of Information Science, Cornell University\\
    severin.engelmann@cornell.edu, susser@cornell.edu
}

\begin{document}

\maketitle

\begin{center}
\small\textit{Accepted for publication at the 2026 AAAI/ACM Conference on AI, Ethics, and Society (AIES).}
\end{center}

\begin{abstract}
The fair AI/ML literature has long recognized two distinct normative issues raised by increasing automation: issues of distributive fairness, on the one hand, and issues of so-called ``representational'' fairness on the other. The former—distributive issues—pertain to challenges automated systems create for the distribution of resources and opportunities across individuals and social groups. The latter—representational issues—refer to challenges automated systems create for the way individuals and social groups are perceived, understood, and accorded social status. While distributive issues have attracted more attention from researchers and advocates, the rise of generative AI---fundamentally expressive systems whose ultimate function is to convey meaning rather than to automate domain-specific decision procedures---has made representational issues increasingly urgent. Yet, as many have pointed out, we lack adequate normative theories for diagnosing and addressing them. These problems are also being discussed within the rapidly developing field of value alignment, particularly through research on what and whose values and perspectives AI systems should represent. These discussions frequently appeal to standards of descriptive accuracy, asking whether generative AI systems accurately depict specific social groups. This strategy, we argue, falls short on several counts. For many (if not most) social groups, there are no stable or bounded referents against which to judge claims of representational inaccuracy. It is unclear who has the authority to decide what counts as misrepresenting any particular group. And even accurate representations can reproduce harmful patterns that are detrimental to group outcomes. The problem, at bottom, is not misrepresentation; the problem, we argue, is misrecognition. Drawing from relevant work in political theory—especially Nancy Fraser’s account of participatory parity—we show how moving from ``representational fairness'' to ``recognitional justice'' equips us with better conceptual and normative tools for understanding and addressing these challenges. 
\end{abstract}

\section{Introduction}
The remarkable expressive capabilities of generative AI systems, including large language models (LLMs), are driving their growing influence throughout society. Users might ask an AI chatbot to explain the ritual practices of a religious holiday celebrated in another country. They might ask it about dating expectations in two different cultures. Or they might ask the model to generate images of a ``traditional'' wedding, a ``safe'' neighborhood, or a ``professional'' woman. Whenever AI systems respond to these kinds of queries they generate value-laden depictions of social groups, cultural symbols, practices, and roles.

Several strands of AI ethics research have raised concerns about this expressive power of generative AI and its implications for individuals and social groups. The fair AI/ML community has long recognized that AI systems can impact how people and groups are perceived, understood, and accorded social status---what that literature refers to as problems of ``representational fairness.'' Researchers in that field traditionally distinguish such issues from problems of ``allocative'' or ``distributive'' fairness---i.e., worries about how AI is used to distribute goods, resources, and opportunities \citep{barocas2023fairness}. Though focused primarily on predictive AI and on questions of distributive fairness, fair AI researchers have begun to develop useful taxonomies for identifying when models characterize social groups in ways that are, for example, stereotypical, demeaning, or toxic (e.g., \citep{corvi2025taxonomizing}). They have recognized the growing salience of ``representational'' harms raised by generative AI, and they have called for deeper theoretical engagement with the normative questions such problems raise.

Questions of representational harm have also become central to other areas of AI ethics, particularly the rapidly developing field of value alignment. Questions about what and whose values and perspectives AI systems should be aligned with are, in important respects, questions about what and whose values and perspectives those systems represent. Pluralistic value alignment investigates whether generative AI models can represent the diversity of values reflected in local, regional, and national norms across the world (e.g., \citep{chiu2025morebench}). For example, descriptive social-science frameworks such as Inglehart and Welzel's World Values Survey have been used to compare model outputs against empirically observed distributions of values across countries and cultures (e.g., \citep{zhang2026cultivating}). Researchers have asked whether generative AI models can preserve minority perspectives rather than collapse them into dominant viewpoints (e.g., \citep{sorensen2024roadmap}). Pluralistic alignment interventions span different stages of the sociotechnical pipeline \citep{engelmann2025llms}. At the data-collection stage, projects such as PRISM seek to collect preference data from heterogeneous populations and enable the development of models responsive to particular social and cultural identities \citep{kirk2024prism}. At the training and evaluation stage, researchers have proposed distributional parity measures that assess whether models produce comparable responses across cultural groups \citep{mushtaq2025towards}, as well as benchmarks that evaluate how closely model outputs approximate pluralistic value datasets such as Value Kaleidoscope \citep{sorensen2024value}.

Lacking a normative account of generative AI's expressive power, research in value alignment often frames the core issue in \textit{epistemic} terms---model outputs are evaluated for truth or accuracy with respesct to a group's self-understanding. Having framed the problem this way, the obvious solution is then to improve AI models' representations of social groups, to make them more accurate and faithful to the preferences, values and cultural symbols associated with different social groups by employing a range of sociotechnical practices, including generating controlled sets of persona prompts to simulate social groups, collecting empirical preference data from members of social groups, and evaluating models' outputs against historical survey data. These approaches, which aim to improve ``representational fairness,'' are intuitive and powerful. False or misleading representations can be deeply harmful, and research exposing how particular models disproportionately misrepresent some groups versus others, and tools for reducing those disparities, are valuable.

However, we argue that such approaches are, at best, incomplete. Evaluating generative AI outputs according to epistemic standards (truth, accuracy) creates a number of its own problems. The boundaries of social groups are often contested, and their self-identity can change and evolve over time, raising difficult questions about how to establish a stable reference against which to judge representational accuracy. At any particular time, there may be disagreement within groups about the accuracy of specific AI outputs, raising questions about who has the authority to decide. And even if everyone agreed that certain texts or images misrepresented a group, there would remain the problem that \textit{accurate} representations can help maintain harmful social hierarchies too.

Instead of framing these issues in epistemic terms, as problems of misrepresentation, we suggest understanding them as fundamentally \textit{political}---problems of ``misrecognition.'' Which is to say, rather than ask if AI depictions are accurate, we should ask if they are \textit{just.} Like fair AI/ML researchers, political theorists have long pointed out that justice demands more than the fair distribution of goods, resources, and opportunities. Justice also pertains to \textit{status}---just societies work to ensure that everyone is recognized as having equal social standing. Viewed through this lens, the problem with generative AI outputs that misrepresent people is not, at bottom, that they are false or inaccurate; the problem is that stereotypical and demeaning representations diminish people's ability to participate in society as equals.

Our aim in what follows is to explain why and how AI ethics communities working on problems of representation should make this conceptual shift. First, we explore the distinction in the existing literature between ``distributive'' and ``representational'' fairness, the overwhelming emphasis on research about the former, and why the rise of generative AI demands greater attention to questions about the latter. Next, we discuss the parallel distinction in political theory between ``justice as redistribution'' and ``justice as recognition,'' providing a high-level (and necessarily selective) overview of debates about what each demands and how they are related, before homing in on one account---Nancy Fraser's ``two-dimensional theory of justice'' and its notion of ``parity of participation''---which, we argue, is especially helpful for parsing questions about the ethics and politics of generative AI. We then turn to the limitations of the ``representational'' frame, demonstrating why epistemic criteria of truth and accuracy can only get us so far. Finally, we show why ``just recognition'' is a more salient normative standard for central ethical challenges in generative AI.

\section{Fair AI's Distributive Focus}

The fair AI/ML literature has long recognized two distinct normative issues raised by increasing automation: issues of distributive (or ``allocative'') fairness, on the one hand, and issues of so-called representational fairness on the other. The former---distributive issues---focus on the role of algorithms in allocating resources, goods, and opportunities, and draws attention to challenges automated systems create for their just distribution across individuals and social groups \citep{barocas2023fairness}. Distributive fairness in AI brought into view a crucial concern about the role of AI models in mediating, structuring, and automating decision procedures that could impact individuals' social, economic, and political position in society. 

The distributive logic assumes that a subpopulation is subject to a domain-specific decision informed by an algorithmic prediction: in hiring, AI prediction estimates a candidate's future job performance or likelihood of success to filter, screen, or rank applicants \citep{kelan2024algorithmic}. In consumer lending, predictive models estimate a borrower's probability of default, which informs loan approval and pricing decisions \citep{addy2024ai}. In criminal justice, recidivism risk models estimate the likelihood that a defendant will re-offend within a specified future time window to inform parole decisions \citep{travaini2022machine}. In medicine, predictive models estimate the probability that a patient has or will develop a particular condition thereby informing treatment decisions \citep{rong2020artificial}. A major achievement of distributive fairness has been the development of a conceptual and technical ``fairness toolbox'' for diagnosing and mitigating harms in predictive AI systems. By introducing formal criteria such as demographic parity and equalized odds, the toolbox made fairness harms specifiable, measurable, and optimizable across specific decision domains \citep{mitchellAlgorithmicFairnessChoices2021a}. 


The second set of issues---described in terms of ``representational fairness''---refers to challenges automated systems create for the way individuals and social groups are characterized, understood, and accorded social status \citep{blodgettLanguageTechnologyPower2020, chien2024beyond}. While the distributive paradigm has been studied extensively, producing definitions that have been operationalized through numerous debiasing tools and fairness libraries \citep{weerts2024can}, research on representational fairness has been a relative backwater. 

One reason for the relative inattention to challenges of representation is the fact that the distributive frame has simply been a better fit for the vast majority of real-world cases researchers and practitioners have worried about in relation to prediction-based automated decision-making---again, in the context of hiring, credit scoring, and so on. A second reason is that while defining, formalizing, and measuring problems of distributive fairness has been difficult enough, doing the same with respect to representational problems is harder still. Questions about distribution can be articulated naturally in quantitative terms, making them somewhat amenable to formalization and computation.\footnote{Of course, in practice it has proven remarkably difficult to formalize and quantify even problems of distributive fairness, as the vast research field that has sprung up around these challenges shows.} Questions about representation, by contrast, are intrinsically qualitative and thus resist from the outset standard computational abstractions.

Though less attention has been devoted to representational problems than to distributive ones, they haven't been ignored entirely. The fair AI community has responded to issues of representation in predictive AI by developing definitions aimed at measuring canonical types of representational harm across different algorithmic systems. In the existing literature, there are at least three different conceptual and definitional approaches: 

First, measuring harm resulting from AI representations has been approached using a (familiar) distributive logic---harmful representations manifest in the aggregate, when some groups are repeatedly and disproportionately associated with attributes, roles, or qualities that can be interpreted as carrying with them certain value judgments. For example, analyzing search engine results for images of particular genders, researchers find women to be depicted as ``smiling'' or ``beautiful'' while men are characterized as ``leaders'' or ``engineers'' \citep{katzman2023taxonomizing}. Such representational harms, identified in the aggregate, apply a demographic parity heuristic: the concern is not whether any particular label-group association is harmful, but whether the distribution of associations across social groups is unequal.

A second approach relies on taxonomies of harm specified at the level of individual evaluative instances that cast a social group in a negative light. Taxonomies of harm serve as deductive lenses for identifying such instances as harmful---identifying stereotypical or demeaning representations, for example, makes it possible to determine when an AI output positions a social group as lesser, deviant, or unworthy of respect \citep{corvi2025taxonomizing}. Here, it is the individual construct attached to a group that is problematic, rather than the distribution of a construct across different groups. A typical example in the literature on representational fairness is the assignment of an animal label to an image of a person of a particular race \citep{wangRepresentationalHarms2022}. Beyond such clear cases of racism, using the deductive lens of a taxonomy to classify individual instances of representational harm necessarily requires navigating the contested nature of cultural semiotics and semantics. 

Another type of representational harm involves the \textit{absence} of representation \citep{qadri2025risks}. Omission disempowers social groups through representational neglect. It can occur in at least two ways. The first is a failure to represent a social group at all, treating the group and its identity-constituting attributes as invisible or nonexistent. The second is erasure: a failure to represent the distinctiveness of a social group by assigning it the symbolic representations of another, typically dominant, cultural group. Over time, the group's distinctiveness dissolves into the dominant group's symbolic repertoire, denying it the agency—and thus the power—to define itself. A common example of erasure in the fair AI/ML literature is the lack of image labels for nonbinary gender identities, which contributes to their marginalization \citep{you2024beyond}.

One can see from these examples that the fair AI community has made some progress on the problem of representation, which we build on in what follows. Still, its dominant focus has been on questions of fair distribution. Representational harms became visible especially in cases where predictive or classificatory systems produced expressive outputs, including image tagging and pre-LLM natural language processing applications, such as machine translation \citep{katzman2023taxonomizing}. But these concerns remained secondary to the field's central focus on predictive systems used to distribute resources and opportunities. That focus was justified---until recently, most high-stakes AI applications involved prediction and classification systems that shaped access to jobs, housing, financial credit, insurance, and other social goods. Real-world cases of AI-driven representational harm were far less prevalent. Now, however, generative AI is fast rebalancing this equation.

\section{Generative AI Shifts the Fairness Discourse}

With the rise of generative models, representational problems are becoming a more central normative challenge for AI ethics. Unlike predictive systems, LLMs and related technologies are fundamentally expressive systems—their primary function is to convey meaning in language, imagery, audio, and video that participates in the cultural construction of social groups \citep{haghighi2025ontologies}. Such constructions are fundamentally evaluative---they carry normative judgments and, in doing so, may invite claims of harmful misrepresentation. In contrast to predictive AI, a single generative model, such as an LLM chatbot, can produce numerous types of evaluations of any imaginable social group. 

First, to personalize interactions, generative AI represents users by inferring demographic data, intentions and goals, moral commitments, psychographic and behavioral profiles, among other attributes \citep{teeny2026promise, anthis2025llm, sorensen2025value}. As the context of interaction shifts, so too may the categories through which users are defined and represented: as voters, patients, workers, parents, tenants, believers, migrants, fans, consumers, students, or members of particular cultural and political communities. A single generative model can represent an individual across the many social groups to which they belong. 

Second, generative AI produces representations of social groups across a wide range of contexts. Models may describe cultural communities in educational settings \citep{mushtaq2025towards}, generate health advice through the lens of specific religious beliefs \citep{shetty2025vital}, portray national, regional, and local value orientations on political issues \citep{yao2025value}, or simulate the judgments of demographic groups---for example, what members of different groups might consider polite, offensive, safe, appropriate, or desirable \citep{sun2025sociodemographic}. To evaluate the accuracy of these representations, researchers either collect preference data directly from the groups in question or draw on existing surveys, polls, and behavioral datasets that measure the relevant attitudes or judgments. These empirical distributions are then treated as ground truth for how generative AI should represent social groups.

Third, the ability to depict social groups using different stylistic frames enables generative AI models to impersonate their members through simulated identities known as ``personas.'' To create intimate companions, friends, romantic partners, or mentors, users can prompt generative AI chatbots to interact with them as members of particular social groups. LLMs can be conditioned to generate content using the conceptual and stylistic features---including vocabulary, structure, and values---that the model associates with a particular social group. In this way, personas turn group representations into interactive stand-ins: they make simulated members of a social group available for dialogue across contexts and modalities \citep{amin2026creating}. Cultural-alignment research, for example, uses persona role-playing to instantiate culturally ``competent'' helper models that rate how well candidate responses represent a target culture, thereby supporting the development of culturally aligned generative AI models \citep{yao2025caredio}.

In contrast with generative AI, predictive systems typically represent individuals within contextually-bounded, task-specific frames, such as creditworthiness in credit-scoring models, employability in pre-employment screening models, or risk of reoffending in recidivism models. As generative AI systems proliferate, the concerns long identified by research on representational fairness---biased associations, harmful group portrayals, group omission and erasure---are no longer confined to domain-specific outputs or fixed labels. They extend across open-ended, multimodal, and context-sensitive representations of the many social groups to which individuals belong. Generative AI's representational flexibility matters for questions of fair representation because a change in the conceptual or stylistic framing of a group can alter which attributes are foregrounded, whether those attributes are cast as virtues or deficits, and what stance toward the group the description invites. 

As generative AI systems continue to proliferate, questions of representation are becoming increasingly important for fair AI. Yet, as fair AI researchers have noted, we lack robust conceptual and normative tools for diagnosing and addressing these harms \citep{blodgettLanguageTechnologyPower2020, anthis2025llm, rauh2022}. Putting these issues into conversation with theories of ``justice as recognition,'' we argue, offers a promising path forward.

\section{Distribution versus Recognition}

Beginning in the 1990s, political philosophers challenged purely distributive theories of justice, arguing that \textit{recognition} is an additional and distinct dimension of social injustice \citep{young1990justice, benhabib1999, fraser2003, honnethRecognitionJustice2004}. These debates emerged from growing dissatisfaction with prevailing conceptions of justice that treated distribution as their primary, and often exclusive, concern. In industrialized societies marked by mass production and specialization, demands for the just distribution of income, property, healthcare, education, and other material means of social life crystallized around a central question: who should possess what, and why? Much like fairness in AI, justice was typically understood in material terms, as a question of how goods, resources, and opportunities ought to be allocated.

One reason for this turn to recognition may have been political disappointment and disillusionment---dissatisfied with slow and piecemeal progress toward economic redistribution, demands for justice shifted to what were perceived as the more achievable goals of protection against humiliation, denigration, and disrespect \citep{honnethRecognitionJustice2004}. But it's also possible that the turn towards recognition reflected substantive moral progress: powerful social movements, such as movements for civil rights, women's rights, and gay rights, made it difficult to deny that cultural degradation and status subordination are not minor forms of injustice. In multicultural, pluralistic societies questions about social status are central to social justice.

In societies shaped by expanding ``transnational contact'' \citep{benhabib1999}, ``pluralising value horizons'' \citep{fraser2003}, and the ``destabilisation of inherited interpretative schemes'' \citep{fraser2008social}, social group identities and differences along lines of religion, nationality, ethnicity, race, gender, and sexuality became highly politicized. Political struggles increasingly centered not only on material deprivation, but also on status inequalities organized around differences between social groups, a development that the distributive paradigm could not adequately explain and address \citep{young1990justice, taylor2021politics}. The central question became whether members of different social groups could be accorded equal respect and social standing without having to assimilate to dominant cultural norms. Recognition emerged, in this sense, as a second dimension of justice---attuned to the ways in which cultural markers of social difference can be the basis of subordination, exclusion, and unequal participation in social life. Theories of justice as recognition ask: what is required for individuals and groups to feel (and be) recognized as equals, and how should institutions be arranged to promote those conditions?

A central question concerns the kinds of difference for which individuals and groups ought to receive recognition. Two dimensions are commonly distinguished \citep{iser2013recognition}. The first, recognition as respect, is grounded in the universal and equal standing of all persons by virtue of their common humanity. It is associated with the principle that all human beings, regardless of race, gender, sexual orientation, or ethnic, cultural, linguistic, and religious background, are moral equals and therefore equally entitled to moral respect \citep{benhabib1999}. The clearest cases of injustice understood in this register arise when individuals or groups are cast outside the bounds of the fully human---for example, by being degraded as mere objects or lesser beings. Such practices do not merely insult or offend; they deny people the basic dignity and moral standing owed to them as human beings. In response, justice as recognition affirms the equal dignity and common humanity that all people share across difference. This Kantian conception of recognition as respect for humanity typically finds political expression in a politics of universalism \citep{margalit1996decent}.

The second dimension, often termed recognition as esteem, concerns the positive valuation of particular features through which individuals and groups understand themselves. Under recognition as esteem, difference is considered valuable, not harmful. Recognizing others means appreciating the attributes, practices, histories, or ways of life that are distinctive to a particular person or group's self-identity. Claims for esteem seek positive appraisal of differences that, in part, dominant value systems have ignored, devalued, or denied. They aim to celebrate devalued traits and, in doing so, highlight rather than eliminate group differences \citep{fraser2003}. Many contemporary social struggles for recognition emerge in precisely these terms: social groups demand recognition for aspects of their identities that may have been treated, historically, as unworthy of public regard. Recognition as esteem has been associated with communitarianism, multiculturalism, and identity politics---political theories that center the need for just societies to affirm particular cultural identities \citep{taylorMulticulturalism1994}.

At their core, recognition theories highlight relational dimensions of justice and the normative significance of difference. Recognition as respect insists that social differences must not undermine equal moral standing, whereas recognition as esteem requires that particular differences be positively affirmed. In both cases, recognitional injustice concerns differences between groups: one group appears to receive, or fails to receive, something of value relative to another. How, then, is recognition distinct from distribution, which also operates through comparisons between groups? What is the relationship between claims to justice as recognition and claims to justice as distribution?

Some theorists have argued that, within a liberal framework, justice can be formulated in an exclusively distributive grammar \citep{rorty1994hidden}. On this ``consumerist view'' view, nonmaterial goods such as rights, power, opportunity, status, dignity, and self-respect can simply be understood as goods to be distributed among social actors. After all, the concept of distribution presumes a relationship of possession: a person or group either rightfully possesses an object of value or does not. Anything considered valuable can, in principle, be divided, and the resulting shares can be compared according to standards of equality. However, other theorists contend that extending this distributive logic to encompass \textit{nonmaterial} social goods, such as respect and esteem, mistakenly characterizes them as static and divisible goods rather than dynamic and relational social processes \citep{young1990justice}. On this view, the nonmaterial, symbolic, and fundamentally relational nature of recognition intrinsically resists the part-of-the-pie logic of the distributive paradigm.

There are also theories that push to the other end of the spectrum, arguing that all forms of injustice---including distributive injustice---can ultimately reduce to failures of recognition. On such accounts, struggles for economic equality and material redistribution are fundamentally expressions of misrecognition \citep{honnethRecognitionJustice2004}. Access to income, work, and opportunities reflects institutionalized judgments about whose contributions are worthy of recognition. Economic inequality is therefore morally significant because it signals that some individuals or groups are not properly recognized as full and valuable participants in social cooperation. Distribution, on this type of view, is downstream from recognition.

But theories that reduce all forms of injustice to failures of recognition have difficulty explaining certain paradigmatic social problems. For example, some cases of maldistribution seem not to arise from misrecognition but from the regular workings of impersonal economic structures \citep{fraser2003}. Consider profit-driven layoffs, which may result from structural market pressures---such as a company's decision to relocate production---rather than from a loss of recognition for a particular form of labor. If, at bottom, all injustice is understood as misrecognition, harms generated by impersonal economic structures must be explained, ultimately, as failures of recognition. Reducing all forms of injustice to misrecognition therefore requires treating structural arrangements themselves as capable of conferring or withholding recognition.

Obviously, we can't resolve these debates here. Nor have we the space to connect them with the many other closely related discussions in political theory that conceptualize these issues in slightly different terms---e.g., debates about competing varieties of egalitarianism, especially theories of relational egalitarianism \cite{anderson1999}. But we hope to have demonstrated the depth and complexity of the problems at hand. In the next section, we discuss in more detail one approach to reconciling questions of distribution with questions of recognition---Nancy Fraser's ``two-dimensional'' theory of justice. Our aim in doing so is to make these abstract theoretical debates somewhat more concrete, before returning to the specific problems raised by generative AI. We draw on Fraser's theory for two reasons: in our view, it is a highly plausible integrative theory of justice, and it parallels, in many ways, existing discussions about related issues in fair AI. However, the overarching argument we are advancing in this paper doesn't rest on the details of Fraser's account.

\section{A Two-Dimensional Theory of Justice}

While many theorists attempt to reduce recognitional claims to distributive ones, or vice versa, Nancy Fraser argues that both recognition and distribution are basic, co-constitutive components of social justice \citep{fraser2003, fraser2008social}. Fraser's nonreductive account reflects her broader view that modern societies are irreversibly globalized, ethically pluralistic, and deeply contested: they encompass multiple, often incompatible value horizons within which social groups struggle to institutionalize their forms of social life. Justice must therefore be understood as multidimensional too, attending to the complex ways societies are structured, not only by economic arrangements but also by contested patterns of cultural value \citep{fraser2008social}.

According to Fraser, trying to diagnose injustice solely in terms of economic exploitation \textit{or} in terms of status subordination oversimplifies matters \citep{fraser2003}. Both economic materiality \textit{and} cultural valuation impact a person's social standing in distinct and fundamental ways. On Fraser's view, the central demand of justice is equal social standing, or what she calls ``parity of participation.'' ``According to this norm,'' she writes, ``justice requires social arrangements that permit all (adult) members of society to interact with one another as peers''\citep[p.~36]{fraser2003}. There are, she argues, two conditions necessary for parity of participation. One, an objective condition associated with distributive concerns that requires people to have ``sufficient material resources that they have the means and opportunities to interact with others as peers'' \citep[p.~36]{fraser2003}. Two, what Fraser calls an intersubjective condition, associated with justice as recognition, which ``precludes institutionalized norms that systematically depreciate some categories of people and the qualities associated with them'' \citep[p.~36]{fraser2003}. The intersubjective condition refers to institutionalized patterns of cultural value that deny some individuals and groups equal social standing,  either by ``burdening them with excessive ascribed ``difference'' or by failing to acknowledge their distinctiveness'' \citep[p.~36]{fraser2003}. 

Fraser's ``two-dimensional'' conception of justice helps explain what's happening when distributive and recognitional issues interact in complex ways. For example, she points to the case of a black Wall Street banker who finds it difficult to hail a taxi \citep[p.~34]{fraser2003}. It also comprehends when someone is subordinated along both distributive and recognitional dimensions at once---for example, a working class person who is also a member of a ``despised'' sexual minority  may suffer both economic exploitation and cultural devaluation in ways that are neither reducible to, nor simply caused by, the other. Parity of participation requires both dimensions of justice: just distribution and just recognition.

The same holds for gender and race. Women may be disadvantaged both through underpaid labor and through sexually objectifying cultural representations \citep[p.~21]{fraser2003}. Fixing one will not automatically fix the other. Likewise, marginalized racial groups may be overrepresented in low paid work while also being subjected to value schemes that privilege traits associated with whiteness \citep[p.~30]{fraser2003}. In such cases, neither distribution nor recognition alone can explain the injustice or how to respond. What remedy is called for, and in what measure, must be determined in relation to the concrete forms of subordination at issue. It must be tailored to the contextual dimension of the specific harm and might require both redistribution and recognition. 

Fraser's approach also provides tools for distinguishing between morally justified and morally unjustified claims of injustice and demands for redress. Not all claims for recognition are warranted and not all forms of uneven distribution are unjust. According to Fraser, claims to injustice are only warranted where parity of participation is undermined, and proposals for remedying injustice are only acceptable if they won't diminish parity of participation themselves. For example, claims by white supremacist groups that their self-esteem is compromised by having to live alongside ``inferior'' races are unjustified and warrant no remedy, since providing the intersubjective conditions they claim to need would result in the subordination of others \citep{iser2013recognition}.

As an evaluative standard, participatory parity tests claims for recognition or redistribution at two levels. First, at the \textit{inter}group level, inequalities must be eliminated when one group's standing is systematically lowered relative to that of another. For example, marriage laws that exclude same-sex partnerships as illegitimate embody an institutionalized pattern of cultural value that casts some social groups as deficient (heterosexual=good, homosexual=bad) \citep[p.~36]{fraser2003}. The result is an institutionalized category of ``devalued persons'' who are denied the opportunity to participate on a par with other members of society. At the intergroup level, claimants must demonstrate that dominant cultural valuations deny them parity of participation.

Second, at the \textit{intra}group level, claims for recognition must be assessed according to whether the practices for which recognition is sought themselves preserve parity of participation. Fraser avoids an identity politics approach to recognitional justice by refraining from locating all normative authority in the moral standards of the claimant group. Not every justice claim advanced is legitimate. If recognition claims would result in the subordination of some members of the group relative to others (or others outside the group), those claims must be rejected. For example, if a minority religious group sought recognition for practices that undermined women's equal standing, those practices would themselves violate participatory parity. In this way, Fraser's deontological standard avoids relativism: claimants must show not only that dominant cultural norms deny them participatory parity, but also that the specific forms of recognition they seek would not further undermine participatory parity, either within the group or beyond it.

Parity of participation is not a standard that can be applied monologically, through a decision procedure that yields a definite answer. There is no objective marker that signals when parity of participation has been achieved, since the meaning of any such marker is itself open to interpretation and contestation. Instead, judgments about whether existing institutionalized patterns of cultural value impede parity of participation, and whether proposed remedies would alleviate those inequalities, must be worked out discursively and dialogically. Claims to recognition must therefore be assessed through public reasoning and contestation. Participatory design, cultural alignment, and pluralistic alignment already incorporate elements of public involvement into sophisticated sociotechnical approaches. Yet there are limits to the kinds of normative deliberation these approaches can bring to bear on fundamentally political questions. What constitutes just recognition cannot be determined by designers or technical procedures alone. It must ultimately be defined, debated, and contested collectively in public.

\section{From Fair Representation to Just Recognition}

Returning to generative AI, questions of how AI models represent, interpret, and shape how social groups are perceived and understood are addressed across AI ethics communities, including fair AI/ML and value alignment.

Recall that generative models produce value-laden depictions of social groups and their cultural symbols via three main modes of interaction. First, they personalize outputs to individual users based on the social groups to which the models infer, over time, those users belong. Second, during interactions, they produce evaluative descriptions of cultural symbols associated with particular social groups. Third, they impersonate members of social groups through personas, acting as teachers, mentors, health advisors, social companions, and so on.

Across these interactive modes, generative AI can undermine participatory parity in a number of ways already discussed in the fair AI literature. First, it can disproportionately associate some groups with socially valued terms and other groups with socially devalued terms---for example, depicting women as ``nurses'' and men as ``doctors.'' Second, AI outputs can denigrate social groups by stereotyping or demeaning them, including clear cases of hate speech. Third, AI outputs can contribute to group erasure by systematically omitting the cultural symbols that define particular group identities. Recent generative AI research documents precisely such phenomena, including group-based distributional differences in representational outputs, overt stereotyping, toxic or denigrating content, and the omission or erasure of social groups' cultural symbols \citep{guo2024bias}. In all of these cases, there is prima facie reason to worry that generative AI is creating or sustaining patterns of cultural value that relegate some groups to subordinate social status. Or as Katzman et al. put it, they are ``contributing to the reproduction of harmful social hierarchies'' \cite[p. 14280]{katzman2023taxonomizing}.\footnote{Another way in which generative AI may cause representational harm is through disparities in quality of service. For example, voice-recognition systems may have lower accuracy rates for underrepresented languages and culturally marginalized users. In these cases, the ability of disadvantaged groups to participate as equals may be undermined in social, political, and economic contexts increasingly mediated by generative AI models. As generative AI becomes more involved in the distribution of socially important resources, the failure to offer fair access could perpetuate existing disparities and deny parity of participation to disadvantaged groups. Consistent with Fraser's two-dimensional account of justice, this illustrates that recognition and redistribution can rarely be neatly delineated in practice. In concrete cases of injustice, the two dimensions are intertwined and vary in their relative emphasis. Disparities in quality of service may therefore simultaneously restrict access to valuable technological resources and reproduce the diminished social standing of disadvantaged groups. For reference, a rich literature in AI fairness has documented cases of disparate quality of service (e.g., \citep{hutiri2024not}).} 

A more difficult case arises from a prominent class of harms framed as inaccurate representations of social groups. A rapidly growing body of work has identified harms from generative models that portray the cultural symbols of social groups in culturally inaccurate ways \citep{dev2026unified, qadri2023ai, qadri2025risks, ghoshGenerativeAIModels2024}. For example, recent work on text-to-image systems identifies harms such as exoticism and cultural misappropriation, in which cultures are represented through homogenized, misplaced, or otherwise culturally inaccurate details. Research on culturally inclusive LLMs argues that models trained on predominantly Western corpora may privilege Western values and thereby misrepresent particular cultural traditions and practices or homogenize culturally specific perspectives in critical domains such as education, health, and ethical or moral deliberation. Research on sociodemographic persona prompting similarly finds that generative AI can produce stereotypical or caricatured representations of the social groups being impersonated \citep{wang2025large}. Such cases may provide strong evidence of cultural domination in Fraser's sense: groups are represented through interpretative schemes that are alien to their own self-understandings, justifying claims of denial of participatory parity and---consequently---of recognitional injustice.

Representational inaccuracy alone, however, does not demonstrate injustice; what matters is whether the representation undermines the equal standing required for parity of participation. An inaccurate group representation may constitute recognitional injustice when it contributes to status subordination---that is, when it undermines ``social arrangements that permit all (adult) members of society to interact with one another as peers'' \citep[p.~36]{fraser2003}. The inaccurate AI output must contribute to a hierarchy of social status, either between groups---for example, between a majority and a minority---or within a group. If an AI output misrepresents a social group through culturally inaccurate symbols in a way that can reasonably be shown to impede that group's parity of participation, there are grounds for seeking a remedy.

One such remedy against inaccurate depictions of social groups in fair representation and culturally aware alignment has been to try and engineer \textit{more accurate} representations through a distinct sociotechnical pipeline. Given diagnostic evidence of inaccurate representation, optimizing models for greater representational accuracy may appear to offer a morally plausible remedy. Although we cannot here provide a comprehensive overview of the different sociotechnical approaches for producing more accurate representations of cultural symbols in generative AI, their central aim is to make group-specific representations––including preferences, values, and perspectives––more faithful to a target taken to represent the group more accurately. 

This optimization template tends to follow three steps. First, researchers construct datasets intended to capture more accurately the preferences, perspectives, or values associated with particular social groups. These data may be newly collected through surveys and human feedback from relevant populations \citep{kirk2024prism}, or drawn from existing global surveys and opinion datasets, such as those incorporated into GlobalOpinionQA \citep{yao2025no}. Second, models are optimized to reproduce these target representations more faithfully, for example through group-specific or individualized reward models, supervised fine-tuning, or distributional preference-optimization techniques. Third, the resulting models are benchmarked against prior versions or other models to assess how faithfully their outputs reproduce the relevant target representation, including through comparisons with empirical group-level response distributions \citep{shetty2025vital}.

The normative reference point in this pipeline is thus fidelity to an ``improved'' representation of a social group. This reference representation may consist of beliefs, preferences, values, or behaviors and practices taken to characterize the group. Greater representational accuracy may seem like a natural remedy for inaccurate representations in generative AI. Yet improved accuracy alone is not a normative criterion of justice as recognition. The question is not simply how AI models can depict a social group's cultural symbols more faithfully. Rather, it is whether an intervention reduces status subordination and thereby improves the group's relative standing in society. Inaccuracy, as specified in research on representational fairness and value alignment, can serve as a useful \textit{diagnostic} indicator of possible misrecognition, but optimizing for greater accuracy does not by itself remedy recognitional injustice.

In fact, accuracy-oriented remedies can raise problems of their own. At least three such challenges merit closer attention. First, accuracy itself can reinforce status subordination and thus impede parity of participation. This problem was already identified in foundational work on algorithmic bias. For example, Safiya Noble's analysis of commercial image search found that, at the time, searches for images of ``doctors'' returned mostly images of men, while searches for ``nurses'' returned mostly images of women \citep[p. 82-3]{noble2013google}. Those search results were more or less accurate---US doctors were more likely to be men and nurses were more likely to be women. Still, depicting doctors as men and nurses as women arguably reinforces cultural associations that relegate women to lower social and economic status. So there's a case to be made for ``inaccurate'' search results that depict greater gender diversity in those professions. (Though, of course, there are reasonable arguments to be made against such an approach too. The issue would need to be decided collectively.)

Put another way, accurate representations can be harmful when they reflect---and help to perpetuate---unjust social realities.\footnote{Green's critique of formal algorithmic fairness in the context of distributive justice offers a broader methodological parallel \citep{green2022escaping}. Green argues that optimizing an algorithm according to a formally specified criterion may produce improvement according to that criterion while sacrificing other normatively valuable principles and leaving the unjust social structures surrounding the decision intact. The point is that optimization relative to a technical target cannot by itself establish that an intervention is just. In the representational context, accuracy is precisely such a technical target.} Generative AI can exhibit such tendencies when it learns prevailing patterns of unjust representation and reproduces them as statistical regularities. This observation raises deeper normative questions that accuracy-oriented remedies can't answer: Should generative AI accurately represent social reality when that reality is structured by injustice, or should it deliberately produce a more idealized but descriptively inaccurate representation? When does departing from existing social reality constitute a justified challenge to status subordination, and at what point does it become an idealized and counter-productive distortion? These questions illustrate that whether a representation is accurate is not by itself the right normative question. What matters is whether, in any particular case, reflecting or departing from real existing conditions contributes to parity of participation. And of course, that's not the only relevant question---the case for inaccurate but recognitionally reparative generative AI outputs would have to contend with possible trade-offs against other socially important values, such as reliability and trust.

The second problem concerns \textit{authority} over accurate representation: who gets to decide for a group what accurately represents it? The AI ethics community has recognized this problem through a turn toward participatory approaches that involve communities in the sociotechnical development of AI models \citep{delgado2023participatory, birhane2022power}. Drawing on traditions such as participatory design, these approaches involve communities affected by AI representations in defining challenges, constructing novel datasets, and evaluating AI output. 

Participatory governance is a powerful response to the concentration of representational authority in developing AI models. They make visible that representational targets are institutionally produced rather than simply discovered. However, as has been noted by researchers in participatory governance (e.g.,\citep{maas2024beyond}), participation is a spectrum that can range from ``token participation'' to meaningful influence over an AI development pipeline. Everything in between is subject to the participation sought by the entity or institution that develops and deploys AI models. Institutions may retain control over participant selection, the scope of deliberation, and whether participant input ultimately changes the model. These three choices alone raise highly complex questions, such as what defines a `relevant' stakeholder, who gets to define what is `relevant' for a representation, and who should oversee the final decision. Yet this limitation does not make participation irrelevant. Its value lies partly in exposing and contesting the political choices hidden behind claims of representational accuracy. 

The third problem with accuracy goes deeper still. Even if the authority question could be settled, any effort to produce a more accurate representation would still need to specify what the relevant group is, which cultural symbols define it, and how its internal plurality should be represented. What would it mean to represent a culture or social group accurately when its boundaries, meanings, and defining characteristics are internally contested and continuously changing? Recognition theorists remind us that in contemporary societies, social groups are hybridized, internally plural, and politically contested; their boundaries, values, and symbols cannot be fixed without controversy \citep{fraser2003, iser2013recognition, rorty1994hidden, benhabib1999}.

As Fraser and other political philosophers note, the cultural composition of contemporary societies is marked by mass migrations, diasporas, and transnational public spheres, which make it impossible to say where precisely one social group ends and another begins \citep{fraser2003}. Members of such societies inhabit different value communities that are neither internally homogeneous nor clearly bounded. Moreover, social groups affect one another through complex interactions. Seyla Benhabib treats culture and its associated symbols as a polyvocal, multivalent, and intergenerational \textit{conversation}, rather than a stable container of specific beliefs and practices \citep{benhabib1999}. It follows that cultural descriptions are politically and ideologically laden, whereby any ``accurate'' representation embeds an authoritative selection among rival inter- and intragroup accounts \citep{rorty1994hidden}.

The objects of representation––cultural symbols––are themselves a matter of political discourse. Once cultural groups are stabilized as targets for optimization, the very remedy that seeks to bring about a more accurate representation might reproduce the pathologies that recognition theorists warn against \citep{fraser2003, iser2013recognition, rorty1994hidden, benhabib1999}. For example, current approaches to improving AI outputs according to accuracy heuristics identify existing outputs as essentializing and flattening a social group's cultural symbols. This tendency follows from the fact that generative models are statistical predictors whose outputs often privilege the most likely continuation within a probability distribution. As Daniel Mwesigwa argues, when applied to culture, this inevitably privileges dominant, statistically frequent expressions of cultural identity \citep{mwesigwa2025against}.

Optimizing for greater accuracy in representing cultural groups runs the risk of becoming essentialist too, because any authoritative description produced through datasets and accuracy metrics could disguise the intracultural politics that occur within every social group \citep{rorty1994hidden, iser2013recognition, fraser2003}. Even a target constructed through participatory methods might compress internal disagreement into a singular or aggregate group representation. Consider adding participatory elements to reinforcement learning from human feedback (RLHF). One way of distributing representational authority more fairly in RLHF is to make groups of annotators more demographically and geographically diverse. Yet, in conventional RLHF pipelines, annotators' judgments between alternative model outputs are ultimately aggregated into numerical reward signals used to train a reward model. Thus, even more divergent views about what constitutes a ``good'' representation can be reduced to a single average optimization target.\footnote{For a general critique of the reductive treatment of human preferences in RLHF, see \cite{zhi2025beyond}.} 

Such remedies might encourage individuals to conform to group representations determined by powerful group members with whom they don't identify \citep{fraser2003}. Participatory designs may make these disagreements visible and provide procedures through which they can be contested, but it cannot transform a fluid and contested social category into a neutral object of measurement and optimization. Formalization requires decisions about whether disagreement should be averaged, weighted, partitioned, or preserved through multiple models or outputs. Each choice constructs a particular account of the group and may elevate some interpretations while marginalizing others. 


In sum, evaluating generative AI outputs in terms of representational accuracy can be diagnostically useful, helping to flag possible cases of recognitional injustice. And participatory approaches can help redistribute representational authority, incorporate marginalized knowledge, and enable affected communities to contest how representational targets are constructed. But less accuracy is not the deep problem and more accuracy is not the deep solution. As shown, accuracy-oriented remedies can fail in at least these three  ways. They can reflect and perpetuate unjust social realities, concentrate authority over whose interpretation counts as representative, and reify the contested social groups they seek to describe. As Fraser puts it, participatory parity cannot be ``applied monologically, in the manner of a decision procedure [...] it cannot be calculated by an algorithmic metric or method'' \citep[p.~42]{fraser2003}. Genuine parity of participation requires public contestation and reasoning. Participants must argue about whether current institutionalized patterns of cultural value undermine participatory parity and whether proposed alternatives would foster it without introducing new injustices of their own.


\section{Conclusion}



Generative AI is an expressive technology that increasingly influences the way individuals and social groups are depicted, judged, and accorded social status. The fair AI literature has traditionally conceptualized the problems this can cause through the lens of ``representational fairness,'' developing conceptual taxonomies and technical tools for identifying stereotypical, demeaning, and other forms of representational harm. Recent research on value alignment, including adjacent (and emerging) communities of cultural and pluralistic alignment, has advanced constructive approaches designed to mitigate such harms, often by working to correct what are perceived as representational inaccuracies.

While these efforts are a valuable starting point, they only take us so far. Connecting pressing questions of representational harm with related discussions in political theory, we've argued that the normative standard to which generative AI models should be held is not \textit{representational fairness} but \textit{recognitional justice.} The deep, underlying question is not whether generative AI outputs are accurate or inaccurate; it is whether they strengthen or undermine what Nancy Fraser terms ``parity of participation''---the ability of all members of society to participate as equals in social life. Answering that question is a qualitative, interpretive, and ultimately political endeavor, which cannot be computed or decided by experts or engineers behind closed doors. What generative AI means for recognitional justice, for our social and political relationships, we must decide collectively, as equals.

\bibliography{references}

\end{document}